\documentclass[aps,prl,reprint,groupedaddress]{revtex4-1}

\usepackage{graphicx,amsmath,color,amsfonts, amsmath, bm}

\newcommand{\be}{\begin{equation}}
\newcommand{\ee}{\end{equation}}

\begin{document}

\title{Universal prethermal dynamics in Heisenberg ferromagnets}

\author{Saraswat Bhattacharyya, Joaquin F. Rodriguez-Nieva, and Eugene Demler}
\affiliation{Department of Physics, Harvard University, Cambridge, MA 02138, USA}


\begin{abstract}

  We study the universal far from equilibrium dynamics of magnons in Heisenberg ferromagnets. We show that such systems exhibit universal scaling in momentum and time of the quasiparticle distribution function, with the universal exponents distinct from those recently observed in Bose-Einstein condensates. This new universality class originates from the SU(2) symmetry of the Hamiltonian, which leads to a strong momentum-dependent magnon-magnon scattering amplitude. We compute the universal exponents using the Boltzmann kinetic equation and incoherent initial conditions that can be realized with microwave pumping of magnons. We compare our numerical results with analytic estimates of the scaling exponents and demonstrate the robustness of the scaling to variations in the initial conditions. Our predictions can be tested in quench experiments of spin systems in optical lattices and pump-probe experiments in ferromagnetic insulators such as yttrium iron garnet. 
  
\end{abstract}



\maketitle

\begin{figure}[b]
\includegraphics[scale = 1.0]{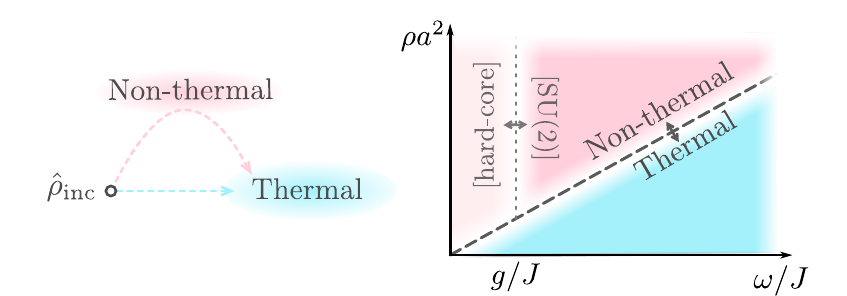}
\caption{The presence of a non-thermal fixed point in phase space can induce a long-lived prethermal state. Appearance of the prethermal state starting from an incoherent state $\hat{\rho}_{\rm inc}$ depends on the average density $\rho$ of magnons pumped into the system and the frequency $\omega$ at which they are pumped, see discussion in the main text. Here $J$ is the exchange coupling, $g$ is the strength of the interaction that breaks SU(2) symmetry and $a$ is the lattice constant.}
\label{fig:schematics}
\end{figure}

{\bf Introduction.} Understanding the emergence of universal dynamics in interacting quantum systems far from thermodynamic equilibrium is a central challenge in theoretical physics. A large body of theoretical works have proposed that isolated quantum systems, such as quark-gluon plasma after heavy ion collision, the early universe after inflation, or cold atoms after a quench, can exhibit universal relaxation of quasiparticles as they evolve far from equilibrium\cite{2003micha,2004micha,2004berges,2011berges1,2008berges,2011berges,2014heavyions,2012yangmills,2014yangmills,2012nonabelianplasma,2014nonabelian,2014kurkela,2015asier}. Such behavior is manifested in universal scaling in time and momenta of correlation functions in the prethermal regime, with the universal exponents independent of microscopic details or initial conditions. The universal relaxation can be attributed to the existence of non-thermal fixed points in the system's phase space (Fig.\ref{fig:schematics}), with the non-equilibrium state inheriting its universal exponents\cite{2015bergesreview}. Manifestations of universal relaxation were recently observed experimentally for the first time in cold atomic gases\cite{2018natureuniversality3,2018natureuniversality1,2018natureuniversality2}. This experimental feat creates new challenges, both in classifying all the possible universality classes as well as devising new tabletop experiments to explore them. 

Here we uncover a new universality class arising in the relaxation dynamics of magnons in the Heisenberg model. Similar to the case observed experimentally with an interacting Bose-Einstein condensate (BEC) \cite{2018natureuniversality1,2018natureuniversality2,2018natureuniversality3}, the prethermal quasiparticle distribution becomes self-similar in momentum ${\bm k}$ and time $t$: 
\be
  n_{\bm k}(t) = t^\alpha f(t^\beta|{\bm k}|),
\label{eq:selfsimilar}
\ee
where $\alpha$ and $\beta$ are universal exponents independent of microscopic details or initial conditions, and $f(x)$ is a universal function. Key properties that determine $\alpha$ and $\beta$ are dimensionality, quasiparticle dispersion and the nature of interactions. While a ferromagnet at low energy can be described as an interacting Bose gas after a Holstein-Primakoff transformation\cite{mattisbook}, the SU(2) symmetry of the Hamiltonian sets the Heisenberg model apart from a conventional BEC in two important ways. First, interaction between quasiparticles are strongly constrained by SU(2) symmetry giving rise to `soft' collisions [Eq.(\ref{eq:interaction}) below]. Second, the SU(2) symmetry suppresses collisions between quasiparticles and the condensate that arises due to symmetry breaking, preventing the renormalization of the quadratic dispersion of quasiparticles [unlike a BEC where Goldstone modes have linear dispersion]. These two features lead to distinct universal exponents in a broad range of wavevectors. 

Crucially, we find that the key requirement for the observation of the universal scaling in Eq.(\ref{eq:selfsimilar}) is that a sufficiently large population of magnons is pumped into the system---the details of the initial conditions are unimportant. In terms of the experimentally controllable density of magnons $\rho$ and pumping frequency $\omega$, the occupation number of modes at frequency $\omega$ is given by $n_\omega \approx \rho a^d (JS/\omega)^{d/2}$, where we assumed that magnons are pumped in an energy window $\Delta\omega \sim \omega$ and have dispersion $\omega_{\bm k} = JS(|{\bm k}|a)^2$ ($a$: lattice constant, $J$: exchange coupling, $d$: system dimension, $S$: spin number). The condition $n_\omega \sim 1$ defines a boundary below which we do not observe self-similarity and the system evolves directly into the thermal fixed point, see Fig.\ref{fig:schematics}. Furthermore, when pumping quasiparticles close to the bottom of the band, we expect interactions that break SU(2) symmetry, {\it e.g.}, dipolar or anisotropic exchange interactions, to be important (Fig.\ref{fig:schematics}). In this case, we expect different universal exponents which will not be discussed here.  

The prethermal state lives for a wide and experimentally accessible temporal window given the slow relaxation of long wavelength Goldstone modes in a ferromagnet. In particular, the thermalization time is determined by terms that break SU(2) symmetry, {\it e.g.}, dipolar interactions. For example, in the context of Bose-Einstein condensation of magnons in yttrium iron garnet (YIG) after microwave pumping\cite{2006demokritov,2007magnoncondensation,2009magnoncondensation}, it was found a thermalization time $\tau \sim 10-100\,{\rm ns}$ that is consistent with dipolar interactions, which are much slower than microscopic timescales associated to the exchange coupling $J\sim 100\,{\rm meV}$. 

Beside of its fundamental appeal, our predictions are relevant to various experiments. For example, we argue that the universal scaling exponents can be accessed in YIG \cite{2006demokritov,2007magnoncondensation,2009magnoncondensation,2014becmagnons,2018magnoninstabilities} given its negligible magnetic anisotropy and despite its ferrimagnetic order\cite{2017yigbands}. In particular, there is a large energy window in which quasiparticles can be pumped such that: (i) the collisions rate due to exchange coupling is much faster than the collision rate due to terms that break SU(2) symmetry and (ii) a single parabolic spin wave mode is populated ($\omega \ll 100\,{\rm meV}$). Cold atom platforms are also promising because the system can be effectively isolated from the environment and the exchange interaction can be engineered using various mechanisms, {\it e.g.} Feshbach resonances, dipolar interactions or lattice shacking\cite{2003amospinexchange,2008amosuperexchange,2013amodipolarexchange,2013amoferromagnet,2014spiralstate,2016hungexchange,2019amospin}.

{\bf Microscopic model.} Focusing only on a single magnon band, we consider a two-dimensional Heisenberg ferromagnet on a square lattice with nearest neighbour exchange and Zeeman field
\be
\hat{\cal H} = -J\sum_{\langle jj'\rangle}{\hat{\bm S}}_j\cdot{\hat{\bm S}}_{j'} + h_z\sum_j\hat{S}_j^z,
\label{eq:Hamiltonian}
\ee
with $J>0$ and $\langle jj'\rangle$ denoting nearest neighbors. We assume that the system has $N$ lattice sites, each containing a spin $S$ degree of freedom, and periodic boundary conditions in each spatial direction. The spin operators satisfy the commutation relations $\left[{\hat S}_j^z,{\hat S}_{j'}^\pm\right] = \pm \delta_{jj'}{\hat S}_j^\pm$ and $\left[{\hat S}_j^+,{\hat S}_{j'}^-\right] = 2 \delta_{jj'} {\hat S}_j^z$, with ${\hat S}_{j}^\pm = {\hat S}_j^x \pm i {\hat S}_j^y$. We also assume that the lattice is at a small temperature such that magnon-phonon interactions can be neglected, which is the case for the prethermal timescales of interest $t\lesssim 1\,\mu{\rm s}$ \cite{mpcoupling1,mpcoupling2}. We include a Zeeman field $h_z$, which is present in many relevant experiments and seems to break the SU(2) symmetry, to illustrate that $h_z$ has no effect on the relaxation dynamics. 

We proceed to build an effective theory valid when the density of quasiparticles is small, $\rho a^2 \ll S$. We recall that one magnon states $|{\bm k}\rangle = \hat{S}_{\bm k}^+|{\rm F}\rangle$ are exact eigenstates of $\hat{\cal H}$ with energies 
\be 
\varepsilon_{\bm k} = h_z + JS(\gamma_0 - \gamma_{\bm k}), \quad \gamma_{\bm k} = \sum_\tau e^{i{\bm k}\cdot {\boldsymbol\tau}}. 
\label{eq:energy} 
\ee
Here $|{\rm F}\rangle = |\downarrow\downarrow\ldots\downarrow\rangle$ denotes the ferromagnetic ground state and $\hat{S}_{\bm k}^+$ denotes $\hat{S}_{\bm k}^+ = \frac{1}{\sqrt{N}}\sum_j e^{-i{\bm k}\cdot {\bm r}_{j}}\hat{S}_{j}^+$. Two magnon states $|{\bm k},{\bm p}\rangle = \frac{1}{2S}\hat{S}_{\bm k}^+\hat{S}_{\bm p}^+|{\rm F}\rangle$, however, are {\it not} eigenstates of $\hat{\cal H}$\cite{1956dyson,mattisbook}. The interaction between magnons can be obtained from the matrix elements 
\be
\begin{array}{c}
\displaystyle\hat{\cal H} | {\bm k},{\bm p}\rangle = (\varepsilon_{\bm k}+\varepsilon_{\bm p})|{\bm k},{\bm p}\rangle + \sum_{\bm q}G_{\bm k,\bm p}^{\bm q}|{\bm k + \bm q},{\bm p - \bm q}\rangle,\\
\displaystyle G_{\bm k, \bm p}^{\bm q} = \frac{J}{N}\left( \gamma_{\bm q - \bm p}+\gamma_{\bm q+\bm k}-\gamma_{\bm q}-\gamma_{\bm k+\bm q-\bm p}\right), 
\end{array}
\label{eq:interaction}
\ee
such that one magnon states are coupled via momentum-conserving collision $G_{\bm k,\bm p}^{\bm q}$. When the incoming magnons have long wavelength, the collision term takes the simple form $G_{\bm k,\bm p}^{\bm q}\approx -\frac{Ja^2}{N}({\bm k}\cdot{\bm p})$ for all values of $S$. This characteristic $({\bm k}\cdot{\bm p})$ interaction, which is independent of $h_z$, arises from the global SU(2) symmetry of the exchange coupling and justifies why magnons propagate ballistically when $|{\bm k}| \rightarrow 0$. Hard core collisions may arise if the SU(2) symmetry is broken.

A long-wavelength description that captures the features of the SU(2) symmetric parent Hamiltonian is
\be
\hat{\cal H} = \sum_{\bm k} \varepsilon_{\bm k}{\hat a}_{\bm k}^\dagger {\hat a}_{\bm k} - \frac{Ja^2}{N} \sum_{{\bm k}{\bm p}{\bm q}} ({\bm k}\cdot{\bm p}) {\hat a}_{{\bm p}+{\bm q}}^\dagger {\hat a}_{{\bm k}-{\bm q}}^\dagger {\hat a}_{{\bm p}}{\hat a}_{{\bm k}}+h.c.,
\label{eq:Hhp}
\ee
where $\hat{a}_{\bm k}$ is a bosonic operator defined after a Holstein-Primakoff transformation of the spin operators. In contrast to the Bogoliubov theory of weakly interacting Bose particles, Eq.(\ref{eq:Hhp}) explicitly shows the absence of anomalous terms ($a_{-\bm k}^\dagger a_{\bm k}^\dagger$) which describe coherent scattering between finite energy quasiparticles and the condensate. Equation\,(\ref{eq:Hhp}) also shows that scattering of quasiparticles vanishes as their momentum approaches zero, which precludes the formation of a magnon condensate at $\bm k = 0$.

{\bf Initial conditions.} We consider an initial incoherent population of magnons in a narrow band of energies centered around $\omega$. Such initial condition can be achieved via transverse microwave pumping $h_\perp(t)$ at frequency $\Omega = 2\omega$\cite{1996gurevichbook}. The amount of magnons pumped into the systems can be controlled by the strength of $h_\perp$, giving two independent knobs to control which ${\bm k}$ modes are excited and their respective population $n_{\bm k}$. Although parametric pumping of magnons also create anomalous correlations $\langle \hat{a}_{-\bm k}^\dagger \hat{a}_{\bm k}^\dagger\rangle $, these decohere rapidly since pairs with different wavevectors $\bm k$ oscillate with different frequencies. We also note that this protocol leads to no net spin texture, $\langle \hat{a}_{\bm k} \rangle =0 $. We parametrize the initial condition as 
\be
n_{\bm k}(t=0) = n_*{\rm exp}\left[-\frac{(|{\bm k}|-k_*)^2}{\Gamma^2}\right], 
\label{eq:initial}
\ee 
where $k_*$ is the wavevector at which magnons are pumped ($\omega = k_*^2/2m$), $n_*$ parametrizes the occupation number of magnons at $k_*$, and $\Gamma$ determines the initial width of the distribution (its value depends on the details of the pump pulse, {\it e.g.}, its duration). We note that, although $n_{*}$ can be much larger than 1, $n_*$ and $\Gamma$ need to satisfy $\rho a^2 \approx k_*\Gamma a^2 n_* \ll S$ for Eq.(\ref{eq:Hhp}) to be valid. 

{\bf Kinetic equation.} The measurable quantity of interest is the magnon population $n_{\bm k}(t) = \langle \hat{a}_{\bm k}^\dagger(t) \hat{a}_{\bm k}(t) \rangle$ as a function of time. In ferromagnetic materials, such quantity can be measured via Brillouin scattering\cite{2007magnoncondensation,2009magnoncondensation}. An alternative technique is spin qubit magnetometry\cite{2017quantumsensingreview}, which has been used to measure (steady-state) magnon population\cite{2015nvmagnons,2017chunhui} as well as imaging single spins\cite{2013singlespinimaging}, but also has been proposed to access a variety of elementary excitations in ferromagnets\cite{2018prlyaroslav,2018nv-ferro}, spin ice\cite{2018spinice}, spin chains\cite{2018nv-wire}, and spin liquids\cite{2018nv-sl}. In cold atom experiments, it is possible to use snapshots of local spin measurements $\langle \hat{S}_i^x\hat{S}_j^x\rangle $ for the different spin pairs in order to compute $n_{\bm k}$.

The time evolution of $n_{\bm k}(t)$ at intermediate timescales (after decoherence of anomalous terms has occured) can be described using the kinetic equation $\partial_t n_{\bm k} = {I}_{\bm k}$, with ${I}_{\bm k}$  the collision integral: 
\be
\begin{split}
{I}_{\bm k} = J^2 a^8 \int_{\bm p}\int_{\bm q}  ({\bm k}\cdot{\bm p})^2 \Big[ n_{\bm k} n_{\bm p} (1+n_{\bm k+\bm q})(1+n_{\bm p-\bm q}) \\ -(1+n_{\bm k})(1+n_{\bm p}) n_{\bm k+\bm q} n_{\bm p-\bm q} \Big]\delta(\varepsilon_i-\varepsilon_f).
\end{split}
\label{eq:collision}
\ee
Here $\varepsilon_i = \varepsilon_{\bm k}+\varepsilon_{\bm p}$ and $\varepsilon_f = \varepsilon_{\bm k + \bm q}+\varepsilon_{\bm p - \bm q}$ the energies of the initial and final states, respectively, and $\int_{\bm p} \equiv \int\frac{d^2{\bm p}}{(2\pi)^2}$. The kinetic equation is valid in the weak coupling regime, $\frac{\int_{\bm k}|I_{\bm k}|}{\int_{\bm k}\omega_{\bm k}n_{\bm k}}\ll 1$, which is true if $\rho a^2 \ll S$. We note that the lack of condensate formation at ${\bm k =0}$ justifies the kinetic equation at long times because, otherwise, condensate formation would give rise to non-perturbative corrections to the kinetic equation. 

\begin{figure}
\includegraphics[scale = 1.0]{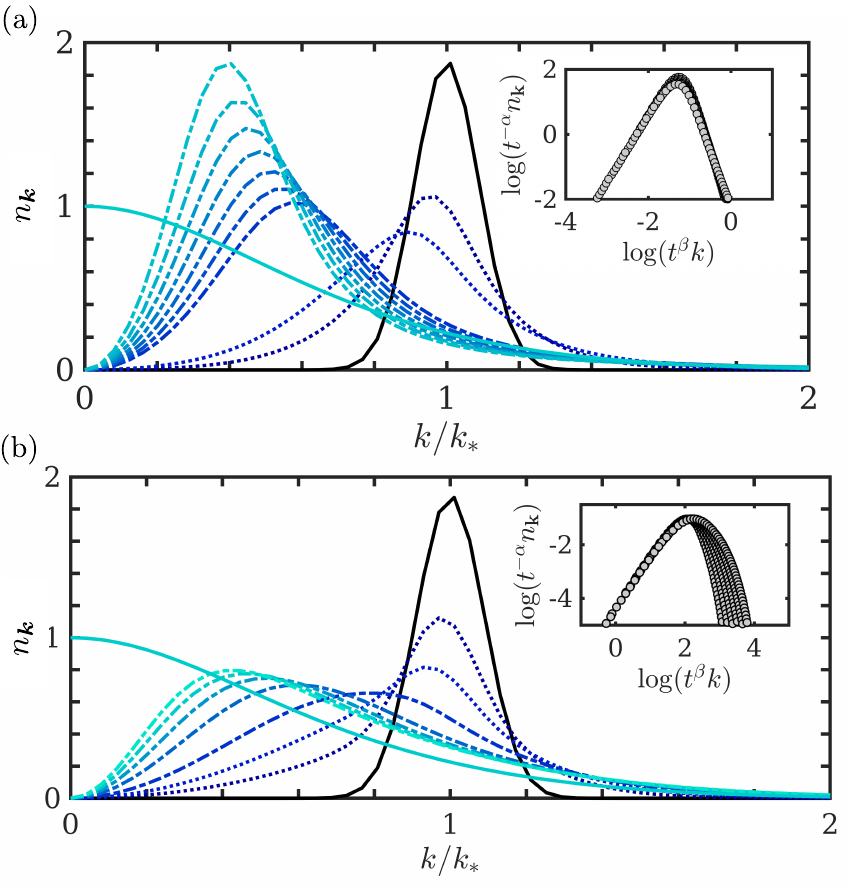}
\caption{Evolution of the occupation number $n_{\bm k}$ starting from an initial incoherent pump at wavevector $|{\bm k}| = k_*$, with occupation (a) $n_* = 100$ and (b) $n_* =1$ [see Eq.(\ref{eq:initial})]. Indicated with dashed-dotted lines is the distribution function once the details of the initial conditions are lost. The inset (a) illustrates the collapse of the data points using a self-similar distribution function in Eq.(\ref{eq:selfsimilar}), with $\alpha$ and $\beta$ defined in Eq.(\ref{eq:parameters}). The inset (b) exhibits no collapse of the data points. Solid lines indicate the initial (black) and final Bose-Einstein (light blue) distribution of the magnon gas. The distributions are plotted at times  $t/\tau_*=0,0.01,0.02,0.09,0.11,0.13,0.16,0.18,0.22,0.28,\infty$ for (a) and $t/\tau_*'=0,0.01,0.02,0.5,0.1,0.15,0.2,0.25,\infty$ for (b), with decreasing tones of blue [$\tau_* = J^2/\omega^3n_*^3$ and $\tau_*' = J^2/\omega^3n_*^2$]. In both panels, we normalized $n_{\bm k}$ with the Bose-Einstein distribution at ${\bm k}=0$ consistent with the energy and particle number of the initial state. Parameters used: $\Gamma = 0.2k_*$, momentum cutoff $\Lambda = 4k_*$(for $n_* = 100$ and $\omega/J = 0.01$, $\rho a^2 \approx 0.1$).}
\label{fig:largedensity}
\end{figure}

\begin{figure*}
  \includegraphics[width=\textwidth]{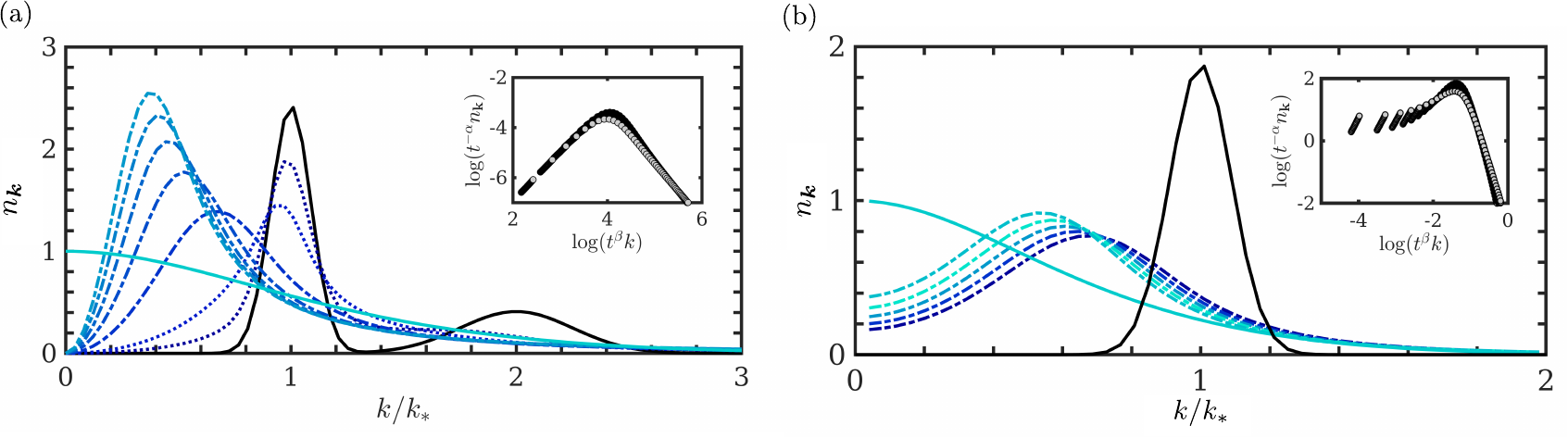}
  \vspace{-10mm}
  \caption{(a) Robustness of the self-similarity under different initial conditions. Shown is the evolution of the distribution function using a two-peak initial condition. After a short time, the distribution function becomes self-similiar like in Fig.\ref{fig:largedensity}. (b) Thermalization in the presence of SU(2) symmetry breaking terms which gives rise to scattering at small momenta. We do not observe a noticeable change in the $(\alpha,\beta)$ values from those found in Eq.(\ref{eq:parameters}). Parameters used are described in the main text.} 
\label{fig:initialconditions}
\end{figure*}

{\bf Universal exponents from kinetic simulations.} We first focus on prethermalization in the large pumping regime, $n_* \gg 1$, where the collision term (\ref{eq:collision}) defines a timescale $\tau_* = \frac{J^2}{\omega^3 n_*^3}$. As shown in Fig.\ref{fig:largedensity}(a), on a timescale ${\cal O}(\tau_*)$, the details of the initial conditions are lost and the distribution function at intermediate momenta, $|{\bm k}| \sim k_*$, acquires a self-similar form governed by Eq.(\ref{eq:selfsimilar}).  We find that the distribution function $n_{\bm k}$ can be fitted by Eq.(\ref{eq:selfsimilar}) in a broad range of momenta (between one and two decades) using the parameters
\be
  \alpha = 0.65\pm0.05 ,\quad\beta = 0.30\pm0.05,
  \label{eq:parameters}
\ee
and the universal function $f(x) \sim 1/x^{2.3}$. The uncertainty in Eq.(\ref{eq:parameters}) is obtained by initializing the simulation from qualitatively distinct initial conditions (but same energy and particle number) and computing the variations in $(\alpha,\beta)$. Figure\,\ref{fig:largedensity}(a) also shows the lack of scattering of magnon states at $|{\bm k}| \approx 0$. For small momenta, the distribution function evolves as $\partial_t n_{\bm k} \sim t |{\bm k}|^2$. Such behavior is cutoff by interactions that break SU(2) symmetry. We emphasize that the exponents $(\alpha,\beta)$ in Eq.(\ref{eq:parameters}) are different from those found in cold atom experiments\cite{2018natureuniversality1,2018natureuniversality2,2018natureuniversality3}, a consequence of different scattering dynamics and different dimensionality (for instance, $\alpha\approx\beta\approx 0.1$ in a 1D Bose-Einstein condensate in Ref.[\onlinecite{2018natureuniversality1}]). 

As shown in Fig.\ref{fig:largedensity}(b), we do not observe self-similar scaling in our numerical results at small occupation number ($n_* = 1$). In particular, we observe that the distribution at intermediate/large momenta relax to the thermal form $n_{\bm k} \propto e^{-\varepsilon_{\bm k}/T}$ without exhibiting self-similarity (small momenta states, $|{\bm k}|\ll k_*$, still scales as $n_{\bm k}\propto t|{\bm k}|^2$), in agreement with previous results on different models\cite{2008berges,2015asier}.

Importantly, the universal exponents are independent of the details of the initial condition. Figure\,\ref{fig:initialconditions}(a) shows the evolution of the quasiparticle distribution after an initial pump at two frequencies. Similarly to Fig.\ref{fig:largedensity}, the details of the initial conditions are lost in a time scale ${\cal O}(\tau_*)$ and the system evolves in a self-similar fashion with the same universal exponents in Eq.(\ref{eq:parameters}).

{\bf Universal exponents from dimensional analysis.} The scaling exponents can be analytically estimated using arguments based in wave turbulence\cite{booknazarenko,bookzakharov}. In the presence of two positive conserved quantities (particle number and energy), a dual cascade typically occurs in two different regions of {\bf k}-space, one with uniform flux of energy and the other with uniform flux of particles. If we assume that a particle cascade dominates at wavevectors $|{\bm k}|\sim k_*$, then magnon number conservation, $ \rho = \int \frac{d^d {\bm k}}{(2\pi)^d} n_{\bm k}(t) = t^{\alpha - \beta d} \int d^d {\bm\kappa} f(|{\bm\kappa}|)$, implies $\alpha =d\beta$ (${\boldsymbol\kappa} = t^\beta {\bm k}$). Alternatively, if an energy cascade dominates at $|{\bm k}|\sim k_*$, then energy conservation, $ E = \int \frac{d^d {\bm k}}{(2\pi)^d} \varepsilon_{\bm k}n_{\bm k}(t) = t^{\alpha - \beta (d+2)} \int d^d {\bm\kappa} f(|{\bm\kappa}|)$, implies $\alpha = \beta(d+2)$. A second relation between $(\alpha,\beta)$ is obtained from the kinetic equations assuming that $n_{\bm k} \gg 1$: the collision term (\ref{eq:collision}) can be rescaled as $I_{\bm k} = t^{3 \alpha - 4 \beta - 2 d \beta +2 \beta} I_{\boldsymbol\kappa}$, and $\partial_t n_{\bm k}$ can be rescaled as $\partial_t n_{\bm k} = t^{\alpha-1} [ \alpha f(|{\boldsymbol\kappa}|) + \beta \kappa f'(|{\boldsymbol\kappa}|)]$.  Matching the coefficients of $t$ of these two terms results in $2(d+1) \beta -2 \alpha  = 1$.

The universal exponents found numerically [Eq.(\ref{eq:parameters})] are modestly close to those associated with a particle cascade ($\alpha_{\rm p} = 1,\beta_{\rm p} = 0.5$); however, we do not see in our simulations a sharp separation between energy and particle cascade. The lack of a dual cascade appears to be linked to a combination of two effects: (i) absence of an emergent lengthscale in Eq.(\ref{eq:Hhp}) ({\it i.e.}, the only lengthscale in the problem is $k_*$ defined in the initial condition) and (ii) the arrested dynamics at intermediate $|\bm k|$ due to absence of scattering at $\bm k \rightarrow 0$. This is distinct from a weakly interacting Bose gas where there is an interaction lengthscale $1/\lambda = \sqrt{mg\rho a^2}$ (which defines a boundary between energy and particle cascades\cite{2015asier}; $m$: mass] and the unconstrained particle cascade towards $\bm k\rightarrow 0$. Incorporating real space dynamics to understand the emergent lengthscales in the relaxation of spin models is a future direction to explore. 

{\bf SU(2) symmetry breaking terms.} Interactions that break SU(2) symmetry allow to populate the ${\bm k} \approx 0$ modes and give rise to different universal exponents if they dominate over Heisenberg exchange. One common interaction is exchange anisotropy, $\hat{\cal H}_{z}= \delta J_z \sum_{\langle i,j \rangle} \hat{S}_i^z\hat{S}_j^z$, which reduces the symmetry of the Hamiltonian from SU(2) to U(1) and effectively gives rise to hard-core collisions $\hat{\cal H}_{z} = \frac{\delta J_z}{N}\sum_{\bm k, \bm p, \bm q}\hat{a}_{\bm k+\bm q}^\dagger \hat{a}_{\bm p - \bm q}^\dagger \hat{a}_{\bm k}\hat{a}_{\bm p}$. Dipolar interaction, in addition, breaks the remaining U(1) symmetry and gives rise to anomalous terms in the Hamiltonian. So long as quasiparticles are pumped at energy scales in which Heisenberg exchange dominate over interactions that break SU(2) symmetry, we expect the exponents $(\alpha,\beta)$ in Eq.(\ref{eq:parameters}) to hold.

To confirm that this is indeed the case, Fig.\ref{fig:initialconditions}(b) shows the magnon relaxation in the presence of a weak exchange anisotropy $\delta J_z = 0.05J(k_*a)^2$ and $n_*\gg 1$. Contrary to the results above, the ${\bm k}\approx 0$ modes are populated. However, the universal exponents at intermediate momenta, where Heisenberg exchange dominates, remain within the values found in Eq.(\ref{eq:parameters}).

{\bf Summary \& outlook.} We showed that the Heisenberg model hosts universal quasiparticle relaxation after an incoherent pump. One direction to explore is whether other universal regimes are possible under different classes of initial conditions, {\it e.g.}, spin textures \cite{magnetictexture2,magnetictexture3} or in the presence of orbital degrees of freedom\cite{2016spinorbec}. Other open questions are whether the same self-similar scaling survives in the large magnon density regime, {\it i.e.} as we approach criticality, and its connection with coarsening dynamics and ageing\cite{1994Bray,2005ageingcalabrese}. Such studies need to go beyond the kinetic equation, for instance, using Truncated Wigner Approximation for spin systems\cite{2015PRX-babadi,spintwa1,spintwa2,spintwa3}. On the experimental front, probing the predicted non-thermal fixed point in ferromagnets is within grasp of ongoing experiments, namely driven ferromagnetic insulators and quenches of spins in optical lattices. 


{\bf Acknowledgements.} We thank C. Du, G. Falkovich, B. Halperin, J. Marino, A. Pi\~{n}eiro Orioli, D. Podolsky, A.~M. Rey, A.~A. Rosch, D. Sels, A. Yacoby, T. Zhou for enlightening discussions. We acknowledge support from Harvard-MIT CUA, NSF Grant No. DMR-1308435 and AFOSR-MURI: Photonic Quantum Matter (award FA95501610323).

\end{document}